\documentclass[12pt]{article}
\usepackage{graphicx}
\usepackage{amssymb}
\usepackage{epstopdf}
\usepackage{multirow}
\newcommand\mathC{\mkern1mu\raise2.2pt\hbox{$\scriptscriptstyle|$}
        {\mkern-7mu\rm C}}              
\newcommand{\mathR}{{\rm I\! R}}         

\newcommand{\be}{\begin{equation}}
\newcommand{\ee}{\end{equation}}

\let\ssection=\section
\renewcommand{\section}{\setcounter{equation}{0}\ssection}

\newcommand\bi{\begin{itemize}}
\newcommand\ei{\end{itemize}}

\begin{document}
\begin{center}
{\large\bf Assessing the Montevideo Interpretation of Quantum Mechanics}
\end{center}

\begin{center}
Jeremy Butterfield\\
Trinity College, Cambridge CB2 1TQ: jb56@cam.ac.uk
\end{center}

\begin{center}
Published in {\em Studies in the History and Philosophy of Modern Physics}, May 2014 (online: DOI: 10.1016/j.shpsb.2014.04.001)
\end{center}

\begin{center} Sunday 15 June 2014 \end{center}

\begin{abstract}
This paper gives a philosophical assessment of the Montevideo interpretation of quantum 
theory, advocated by Gambini, Pullin and co-authors. This interpretation has the merit of 
linking its proposal about how to solve the measurement problem to the search for quantum 
gravity: namely by suggesting that quantum gravity makes for fundamental limitations on the 
accuracy of clocks, which imply a type of decoherence that `collapses the wave-packet'.

I begin (Section 2) by sketching the topics of decoherence, and quantum clocks, on which 
the interpretation depends. Then I expound the interpretation, from a philosopher's 
perspective (Sections 3, 4 and 5). Finally, in Section 6, I argue that the interpretation, 
at least as developed so far, is best seen as a form of the Everett interpretation: namely 
with an effective or approximate branching, that is induced by  environmental decoherence 
of the familiar kind, and by the Montevideans' `temporal decoherence'.

\end{abstract}

\begin{center}
Keywords:  Montevideo interpretation; quantum clocks; decoherence; Everett interpretation
\end{center}

\newpage
\tableofcontents

\newpage

\section{Introduction}\label{prosp}
This paper gives a philosophical assessment of a recent proposed interpretation of quantum 
theory, advocated by Gambini, Pullin and co-authors, and called by them `the Montevideo 
interpretation'. (So I will dub these authors `the Montevideans'.) Although the 
interpretation  is
bound to be controversial, it has the merit of linking its proposed solution to the 
measurement problem to the search for quantum gravity: in short, by suggesting that quantum 
gravity makes for fundamental limitations on the accuracy of clocks, which imply a specific 
temporal type of decoherence that `collapses the wave-packet'. For it is surely a merit to 
link debate about quantum foundations to the search for new physics, even speculative new 
physics.

I therefore begin by sketching the standard topics on which the interpretation depends 
(Section \ref{scape}). Then
I expound the interpretation, from a philosopher's perspective (Sections 3, 4 and 
5).\footnote{This interpretation has been developed in about a dozen papers over the last 
ten years. As I see matters, the main ones are: Gambini,  Garcia Pintos and Pullin (2010, 
2011, 2011a),
Gambini, Porto and Pullin (2006, 2007, 2008), Gambini and Pullin (2007, 2009, 2009a). For the way in which considerations of quantum gravity, especially the `problem of time'  in quantized general relativity, have motivated the interpretation, cf.  Gambini, Porto, Pullin and Torterolo 
(2009) and Gambini and Pullin (2009a). But my summary will draw on just a few aspects of Gambini, Porto and Pullin (2007), and Gambini 
and Pullin (2009, 2009a); where I would recommend a philosopher to begin their reading. The 
papers are also available at a website: http://www.montevideointerpretation.org/.} Finally, 
in Section 6, I argue that the interpretation, at least as developed so far, is best seen 
as a form of the Everett interpretation: namely with an effective or approximate branching, 
that is induced by  environmental decoherence of the familiar kind, and by the 
Montevideans' `temporal decoherence'.

\section{The landscape}\label{scape}
I introduce the measurement problem and quantum clocks, in Sections \ref{Everett} and 
\ref{clocks} respectively. Then we will be ready for a prospectus about the Montevideo 
interpretation (Section \ref{road}). 

\subsection{Collapse, Everett, decoherence---and gravity}\label{Everett}
I will recall the relevant aspects of  the  measurement problem, by briefly sketching: the 
collapse of the wave-packet (Section \ref{CWP}), the Everett approach and decoherence 
(Section \ref{EDP}), and how the problem may be altered by considering gravity (Section 
\ref{grav?}). Later, I will return to these aspects in more detail: the first three in 
Sections \ref{two} and \ref{fifthproposal}, and the last in Section \ref{fourproposals}.

\subsubsection{The collapse of the wave-packet}\label{CWP}
Let us begin with the orthodox---or at least, traditional and minimalist---approach to 
securing that measurements have definite outcomes. `The collapse of the wave-packet' refers 
to an irreducibly indeterministic change in the state of a quantum system, contravening the 
deterministic and continuous evolution prescribed by the Schroedinger equation. Anyone who 
advocates such a collapse, as a {\em bona fide} physical process that occurs to an isolated 
system, faces several questions. Three of the most pressing are as follows. Under exactly 
what conditions does the collapse occur? What determines the physical quantity (the basis) 
with respect to which it occurs? How does it mesh with relativity? Good questions, which 
went largely un-addressed by quantum theory's founding fathers: but which in recent decades 
have been addressed by a great deal of good work. Section \ref{two} will mention one main 
line, the dynamical reduction programme of Ghirardi and others.

\subsubsection{Everett, decoherence and patterns}\label{EDP}
The Everett or many-worlds interpretation proposes to reconcile quantum theory's 
deterministic evolution of the quantum state with the {\em apparent} collapse of the wave 
packet, i.e. with measurements having definite outcomes with various frequencies, by saying 
that measurement processes involve a splitting of the universe into branches. Obviously, 
this proposal has to face its own versions of  the murky questions just mentioned:  about 
the conditions under which a branching occurs, how we should understand branching, and how 
branching can mesh with relativity. Murky indeed. So it is hardly surprising that
this interpretation has traditionally been regarded as vaguer and more controversial than 
others. Thus Bell, in his masterly (1986) introduction to interpreting quantum theory, 
wrote that it `is surely the most bizarre of all [quantum theory's possible 
interpretations]' and seems `an extravagant, and above all extravagantly vague, hypothesis. 
I could almost dismiss it as silly' (pp. 192, 194).

\indent But I submit since Bell wrote, Everettians have made major improvements to their 
interpretation. In my opinion, there have been two main improvements relevant to our 
purposes: which I will label ``Decoherence' and `Patterns'.\footnote{I set aside a third 
improvement, viz.  various arguments justifying, from an Everettian perspective, the 
orthodox (Born-rule) form of quantum probabilities. All three improvements have been 
developed in many papers over the last twenty years. Some of the latest work is in Saunders 
et al. (2010), and by Wallace (2012, 2012a). The first of these also contains penetrating  
critical assessments of all three improvements by non-Everettians.}

\paragraph{Decoherence}\label{2121}:\\
Although the fundamental ideas of decoherence were established in the early years (and were 
clear to maestros such as Heisenberg, Mott and Bohm), detailed models were only developed 
from about 1980; (Schlosshauer (2008) is an excellent recent survey).

Recall that 'decoherence' means, broadly speaking, the diffusion of quantum coherence from 
the system to its environment. This is the fast and ubiquitous process whereby, for 
appropriate physical quantities, the interference terms in probability distributions, that 
are characteristic of the difference between a superposition and a mixture, diffuse from 
the system to its environment. In a bit more detail: at the end of the decoherence process, 
the quantum probabilities for any quantity on the system are as if the system is in one or 
other of a definite set of states. In many models of how a system (such as a dust-particle) 
interacts with its environment (such as air molecules), this set consists of coherent 
states: states which are sharply peaked for both position and momentum, so that a system in 
any such state is presumably nearly definite in both position and momentum. (But the 
distributions have enough spread so as to obey the Uncertainty Principle's veto on 
simultaneous precise values for position and momentum.)

For our purposes, decoherence has two important features. The first is a kind of 
imprecision. That sounds like a defect; but I shall maintain---especially later, in 
Sections \ref{two} and \ref{enfin}---that this imprecision is, for the Everettian, a merit. 
(Here I follow in the footsteps of some avowed Everettians, such as Saunders and Wallace.) 
So I will call this feature `flexibility'.

Thus we expect the classical physical description of the world to be vindicated by quantum 
theory---but only approximately. Only some subset of quantities should have definite 
values. And maybe that subset should only be specified contextually, even vaguely. And 
maybe the values should only be definite within some margins of error, even vague ones. 
Decoherence secures this sort of flexibility. For the selection of the quantity that is 
preferred in the sense of having definite values (relative to a branch) is made by a 
dynamical process---whose definition can be legitimately varied in several ways. Three 
examples: the definitions of the system-environment boundary, and of the time at which the 
interaction ends, and of what counts as a state being 'sharply peaked' for a quantity, can 
all be varied.

The second feature is that decoherence does not just by itself solve the measurement 
problem. More precisely: it does not imply that the system is in one of the set of states 
(typically coherent states). It implies only that the quantum probabilities are as if the 
system were in one. Furthermore, the theory implies that the system is in fact not in one 
of those states (on pain of contradicting the original hypothesis that the total 
system-plus-environment is in a superposition, not a mixture). This feature is well-known, 
and has been given various names, especially `the problem of outcomes', `the problem of 
improper mixtures' (following a jargon of D'Espagnat) and the `the problem of replacing 
`and' by `or' (following a jargon of Bell); (cf. my discussion of (Outcome) in Section 
\ref{two}).

To put this feature vividly, in terms of Schroedinger's cat: at the end of the decoherence 
process, the quantum state still describes two cats, one alive and one dead. It is just 
that the two cats are correlated with very different microscopic states of the surrounding 
air molecules. For example: an air molecule will bounce off a wagging upright tail, and a 
stationary downward one, in different directions! Since one's overall aim is to solve the 
measurement problem, this feature is usually considered a defect, not a merit, of 
decoherence. But we will now see how the Everettians' second main development may turn it 
into a merit ...

\paragraph{Patterns}\label{2122}:\\
The second development is the application to the problems of quantum ontology, especially 
Schroedinger's cat, of the philosophical idea  that the objects in `higher-level' ontology, 
e.g. a cat, are {\em not} some kind of aggregate (e.g. a mereological fusion) of 
lower-level objects, but are dynamically stable patterns of them, of a special type---which 
type being spelt out by what we believe about objects of that kind. This idea is often 
associated with ``functionalism'' in the philosophy of mind (e.g. Dennett 1991).

Some prominent Everettians, such as Saunders and Wallace, maintain that it snatches victory 
from the jaws of defeat: the defeat, just mentioned, that decoherence apparently does not 
by itself solve the measurement problem. The idea is that the final quantum state's 
describing two cats, one alive and one dead, is a matter of the state encoding two 
patterns---and the description is entirely right.

This becomes a bit clearer if we adopt a specific representation of the quantum state, for 
example position. Then, roughly speaking: the final state is a wave-function on the cat's 
classical configuration space, with two peaks: one peak over some classical configurations 
each corresponding to a live cat, e.g. with a wagging upright tail, the other peak over 
some classical configurations corresponding to a dead cat, e.g. with a stationary downward 
tail. But in that case: according to the idea of cats as patterns, the quantum state does 
indeed represent {\em two} cats.

In other words: we see that we {\em should} take the measurement problem to be solved by 
exactly what decoherence secures: a final state describing a living cat and a dead one. In 
brief: the philosophical idea of higher-level objects as patterns vindicates the Everettian 
vision of a multiplicity of objects.

It is worth stressing (as Wallace, for one, does) that this line of thought is independent 
of quantum theory's details; and so it is also independent of its various weird features 
(e.g. non-locality). The point is closely analogous to one which we all unhesitatingly 
endorse for several other physical theories. Namely, theories in which states can be added 
together to give a sum-state, while the component states are dynamically isolated, or 
nearly so (i.e. do not influence each other). Examples include the theory of water-waves, 
or electromagnetism. So, says Wallace, we should also endorse it in quantum theory, and 
accept that there are two cats.

For example: the water in Montevideo harbour can get into a state which we describe as, 
e.g. a wave passing through the harbour's centre heading due West; or into a state which we 
describe as a wave passing through the centre heading due North; or into a state which is 
the sum of these. But do we face a `Montevideo water paradox'? Do we agonize about how the 
Montevideo harbour water-system can in one place (viz. the harbour's centre) be 
simultaneously both Westward and Northward? Of course not. Rather, we say that waves are 
higher-level objects, patterns in the water-system; and that there are two waves, with the 
contrary properties, one Northward and one Westward. Similarly for the electromagnetic 
field in a certain region, and e.g. pulses of laser light travelling in different 
directions across it.

And similarly, says Wallace, about the quantum wave-functions defined on the classical 
configuration space. There is a state with two cats, one alive and one dead. And of course, 
there are also myriad other states, the vast majority of which do not represent macroscopic 
objects (patterns!) which we might recognize (as cats or as dogs or as puddles or as mud or 
...)---or even sums of these.

\paragraph{Summing up}\label{2123}:\\
Let me sum up these two developments, Decoherence and Patterns. Nowadays, some Everettians 
propose to combine the physics of decoherence with the philosophical (``functionalist'') 
idea that objects  in `higher-level' ontology are
dynamically stable patterns of lower-level objects. This suggests that  the proverbial 
Schroedinger's cat measurement involves an approximate and emergent splitting, after which 
there really are two cats (or two broad kinds of cat), since the total wave-function is 
peaked over two  distinctive kinds of pattern in the classical configuration space: the 
legs, tail and body all horizontal, still and cool (`dead'); and  the legs and tail 
vertical, moving and warm (`alive').

I think these Everettians would admit how hard it is to get one's mind around the central 
idea: the dizzying vision of decoherence processes yielding a continual, but approximate 
and emergent, branching of the universe that: \\
\indent \indent (a): meshes fundamentally with relativity, so that there is no absolute 
time ordering to the splitting; and \\
\indent \indent (b): is to be combined with  almost all objects---not just macroscopic 
objects like cats, tables and stars, but anything that classical physics successfully 
describes as having a spatial trajectory etc.: for example, large molecules---being treated 
as patterns; or better: being treated as the quantum state being peaked above such patterns 
in a classical configuration space.

I said `some Everettians', since I do not think there is unanimity among them. One main 
question is to what extent they should try to make this dizzying vision more precise in 
general terms, perhaps by finding a new mathematical language to describing the emergent 
branching. Thus Deutsch, in the course of his striking appeal (2010) to fellow-Everettians 
to stop defending the interpretation against accusations and
rivals, and instead explore the new physics that it promises to contain, admits that the 
exploration will be very challenging, since no one has yet given a precise mathematical 
descriptions of (even toy models of) this branching structure (2010, p. 546).

On the other hand, Saunders and Wallace (at least as I read them) do not think any such 
precise description could, or should, be possible. Nor is it desirable:  vagueness about 
when branching occurs is a merit---recall the label `flexibility'. They draw an analogy 
which may be helpful: between (a) times as understood in the `block-universe' view of time 
suggested by relativity theory (also known in philosophy as: the `B-theory' of time), and 
(b) worlds or branches as understood by the Everettian. Recall that the advocate of the 
block-universe says:\\
\indent (i): reality is four-dimensional, and slices across it are in principle arbitrary 
and artefactual (especially in relativity theory, with no absolute simultaneity); but 
also\\
\indent (ii): for describing the history of the universe---and in particular, in physics, 
for doing dynamics---only a small subset of slices will be useful; though the criteria to 
select that subset will be a bit approximate.\\
Then the proposed analogy is as follows. Similarly, the Everettian says:\\
\indent (i'): the `slicing' of reality by choosing a basis in Hilbert space is in principle 
arbitrary and artefactual; but also\\
\indent (ii'): for describing the history of the universe---and in physics, for doing the 
approximate and emergent dynamics of a world---only a small subset of bases will be useful; 
though the criteria to select that subset will be a bit approximate (since decoherence 
gives no absolute criterion for a system-environment split, or for when interference terms 
are small enough).

 In short: the analogy is between arbitary slicings of spacetime, and arbitrary choices of 
basis in Hilbert space. Both are precise; in both cases, for doing dynamics, only a small 
subset will be useful; in both cases, the criteria to select that subset will be a bit  
approximate.\footnote{For more details, cf. Wallace (2001); or more briefly, Butterfield 
(2002, Sections 7 and 8).}

So this is very much an open debate among the Everettians.\footnote{And of course, it 
echoes a wider debate about whether fundamental physical theories must be precise. In this 
context, I recommend Kent's (2010) critique of the Everett interpretation.} But I can duck 
out of pursuing it: for my purposes in assessing the Montevideans, it will be enough that 
some Everettians maintain that the vagueness of the branching is a merit.

\subsubsection{Gravity?}\label{grav?}
I will make two remarks about how considerations about gravity might contribute to solving 
the measurement problem. Both are very programmatic and brief, reflecting the fact that we 
do not have a satisfactory unified treatment of quantum theory and gravity; nor even a 
consensus about what it would look like.\footnote{For philosophical introductions of the 
issues involved, cf. e.g. Butterfield and Isham (2001),
and Rovelli (2006).} But they are worth making, to set the stage for the Montevideo 
interpretation.

First: in the controversies about what a quantum theory of gravity should look like, one 
central and long-established theme has been the conflict between the way general relativity 
treats space and time, namely as dynamical (variable, and interacting with matter and 
radiation), and the way other theories, both classical and quantum, treat them, namely as 
non-dynamical (also called: `fixed' or `background'). Besides, in one main approach to 
quantum gravity, viz. the quantization of canonical (i.e. Hamiltonian) formulations of 
general relativity, this conflict is especially striking, indeed severe, as regards the 
treatment of time rather than space: and accordingly, it is called `the problem of time'. 
Though I shall not pursue this problem, I should emphasize that the Montevideo interpretation grew out 
of the Montevideans' engagement with it. For details, cf. e.g. Gambini, Porto, Pullin and Torterolo 
(2009) and Gambini and Pullin (2009a).

Second: there are various proposals that gravity might help solve the measurement problem. 
One example is Penrose's proposal that gravity drives the collapse of the wave-packet; or a 
little more precisely: that a sufficiently large difference in the gravitational 
self-energy of two configurations of masses renders superpositions of such 
configurations unstable. The Montevideans' proposal will be different: the idea will be that quantum 
gravity implies fundamental limitations on clocks. To set the stage, the next Section 
briefly discusses quantum clocks.

\subsection{Quantum clocks}\label{clocks}
The treatment of time in quantum theory is a large and controversial subject---though 
fortunately, not as large and controversial as the measurement problem as a whole! But in 
the last few decades, the situation has become clearer through the work of many people on 
such topics as the time-energy uncertainty principle, and the conditions for the existence 
of a time operator.\footnote{Butterfield (2012) is a philosophical introduction, 
emphasizing the first topic, especially the work of Busch, Hilgevoord and Uffink. For the 
whole subject of time in quantum  theory, Muga et al. (2008) is an excellent collection; 
for a glimpse of work since then, I recommend Yearsley (2011).}

To prepare for the later discussion, I need here only to sketch two themes about quantum 
clocks. Both concern limits to their resolution, arising from uncertainty principles. The 
first theme is elementary and general, and so I can give some details; the second involves 
detailed arguments about how certain ideal clocks might be constructed, and I will just 
cite the results the Montevideans will appeal to.  These themes are: \\
\indent (i): the time-energy uncertainty principle constrains the resolution of clocks, 
however they might be constructed: roughly, narrowness in the energy distribution worsens 
the resolution of a clock;\\
\indent (ii): applying uncertainty principles to certain ideal clocks implies that  mass 
constrains the resolution: roughly, a lower mass worsens the resolution.

(i): {\em The time-energy uncertainty principle}:\\
I shall assume that here, `time' refers to a dynamical variable of the quantum system, 
whose value, especially  expectation value, serves as a clock-pointer, i.e. to measure 
time. Thus in  principle, every non-stationary quantity $A$ defines for any state $\rho$ a 
characteristic time $\tau_{\rho}(A)$ in which $\langle A \rangle$ changes `significantly'. 
For example: if $A = Q$, and $\rho$ is a wave packet, say $\rho = | \psi \rangle \langle 
\psi |$, then $\tau_{\rho}(A)$ could be defined along the lines: the time for the bulk of 
the wave packet to shift by its width. This assumption is not controversial. It corresponds 
to what Busch in his excellent survey (2008) calls `intrinsic time' (and in his earlier 
(1990, 1990a), `dynamical time').

With this notion of time, we can easily derive what is perhaps  the best-known time-energy 
uncertainty principle, the Mandelstam-Tamm principle. Namely, we define the
characteristic time for any quantity $A$ by
\be
\tau_{\rho}(A) \; := \; \Delta_{\rho} A \; / \; | \; d \langle A \rangle_{\rho} \; / dt \; 
|
\label{tauMT}
\ee
and readily deduce (e.g. Messiah 1965, p. 320)
\be
 \tau_{\rho}(A) \Delta_{\rho}(H) \geq  \frac{1}{2} \hbar .
\label{MTUP}
\ee
An easy example is given by a free particle in a pure state $\psi$ with a sharp momentum 
i.e. $\Delta_{\psi}P << | \langle P \rangle_{\psi} |$. We can deduce that $\Delta_{\psi} 
Q(t) \approx \Delta_{\psi} Q(t_0)$, i.e. the wave-packet spreads slowly; so that the 
Mandelstam-Tamm time $\tau_{\rho}(Q)$, as defined by eq \ref{tauMT}, is indeed the time it 
takes for the packet to propagate a distance equal to its standard deviation.

But to discuss the resolution of a quantum clock,  we need instead the example of the 
lifetime of a property represented by a projector $\hat P$. Let us define $p(t) := \langle 
\psi_0 | U^{-1}_t {\hat P} U_t | \psi_0 \rangle$; where of course $U_t := 
\exp(-itH/\hbar)$,  so that $p(t)$ is the probability of ``having'' $\hat P$ at $t$. Then 
the Mandelstam-Tamm uncertainty relation gives
\be
| \frac{dp}{dt} | \leq \frac{2}{\hbar} (\Delta_{\psi_0} H) [p (1 - p)]^{\frac{1}{2}} \; .
\ee
Integration with initial condition $p(0) \equiv 1$ (i.e. $\hat P$ actual at $t = 0$) yields
\be
p(t) \geq \cos^2 (t(\Delta_{\psi_0} H) \; /\; \hbar) \;\; \mbox{for} \;\; 0 < t < 
\frac{\pi}{2}\frac{\hbar}{(\Delta_{\psi_0} H)} \; \; ;
\ee
so that if we define the lifetime $\tau_{\hat P}$ of the property $\hat P$ as a 
``half-life'', i.e. by $p(\tau_{\hat P}) := \frac{1}{2}$, we deduce
\be
\tau_{\hat P}.\Delta_{\psi_0} H \geq \frac{\pi \hbar}{4} \; .
\label{MTlifetime}
\ee
The general point is that the rate of change of a property of the system decreases with 
increasing sharpness of the energy: and in the limit of an energy eigenstate,  all 
quantities have, of course, stationary distributions.

It is straightforward to apply the Mandelstam-Tamm inequality eq. \ref{MTlifetime} to get a 
lower bound on the resolution of a quantum clock. The general  idea of such a clock is that 
it is a sequence $\psi_1, \psi_2,...$ of orthogonal clock-pointer states that are occupied 
at equi-spaced times $t_1, t_2,...$; so the clock resolution is $\delta t := t_{k+1} - 
t_k$. Thus suppose the clock-pointer is  the mean position of a wave packet. Then eq. 
\ref{MTlifetime}, together with the constraint that the resolution be greater than the 
Mandelstam-Tamm time, i.e. $\delta t \geq \tau_{\psi}(Q)$, implies
\be
\delta t \geq \tau_{\psi}(Q) \geq \frac{\pi \hbar}{4}\frac{1}{\Delta_{\psi}H} \; .
\ee
So the sharper the energy, the worse the resolution: more exactly, the higher is the 
resolution's lower bound.\footnote{There is  a great deal more to say here, even without 
going into details about clocks' construction as the arguments cited in (ii) will do. For 
there is a large literature on various versions of the time-energy uncertainty principle 
(Busch 2008): for example, Butterfield (2012) advertises how the Hilgevoord-Uffink 
uncertainty principles have advantages over the usual ones.}

(ii): {\em A constraint from mass}: \\
There are some heuristic arguments that apply uncertainty principles (of position-momentum, 
as well as time-energy) to certain ideal clocks (e.g. bouncing light between two mirrors), 
and conclude that there are fundamental limits on the clocks' resolution: limits additional 
to those sketched in (i) above.

Thus in a famous analysis, Salecker and Wigner (1957) argued that the mass of a clock 
limited its accuracy: specifically,  if $T$ is the time-interval to be measured by a clock 
of mass $M$, then there must be an  uncertainty $\delta T$ proportional to $\surd{T/M}$.
This  analysis suggests that one can attain arbitrary accuracy by a sufficiently massive 
clock. But Ng and van Dam (1994) argued that in general the mass needs to be confined, and 
that therefore the attainable mass, and so accuracy, is limited by the onset of the 
formation of a black hole (!). I need not go into the details. Here it will suffice to say 
that the Montevideans will endorse  this line of argument (Section \ref{heur}).

\subsection{The road ahead}\label{road}
So much by way of sketching the landscape. From now on, I turn to assess the Montevideo 
interpretation of quantum theory. As I said in Section \ref{prosp}, the proposal is to 
solve the measurement problem by combining decoherence (as in Section \ref{Everett}) with a 
fundamental limitation on the accuracy of quantum clocks (as in Section \ref{clocks}) which 
is suggested by some heuristic considerations about quantum gravity.

The plan is as follows. I will first summarize the Montevideans' analysis of how, in 
orthodox quantum theory, inaccurate clocks make for mixed quantum states (Section 
\ref{montev1}). Then I discuss these states' interpretation, by comparison with the mixed 
states obtained in two other approaches, viz. decoherence by the environment and dynamical 
reduction (Section \ref{meaning1}). In Section \ref{montev2}, I report the Montevideans' 
main proposals: that considerations about quantum gravity make for a fundamental limitation 
on the accuracy of quantum clocks, which solves the measurement problem---in effect, by 
answering objections that are made to the mixed states obtained by environment-induced 
decoherence. Finally, in Section \ref{enfin}, I will urge that the Montevideo 
interpretation is best seen as a version of the Everett interpretation, with an effective 
or approximate branching. It will be clear that the discussion becomes more controversial, 
as we proceed: one can `get off the bus' at various stops. That is: in the present state of 
theoretical and experimental knowledge, the later proposals remain unproven---but 
nonetheless interesting!

\section{Evolution with respect to a real clock}\label{montev1}
The Montevideans' first contribution is a natural sequel to Section \ref{clocks}. For it is 
an analysis within orthodox quantum theory of what the phrase, `the probability of a given 
outcome, for a measurement on a given state at a given time', really means, when one takes 
seriously that the `given time' is as read by a quantum quantity. (This Section summarizes 
part of Section II of Gambini, Porto and Pullin (2007), and follows their notation.)

We envisage measuring a quantity $\hat O$ on a quantum system, when a quantity, $\hat T$ 
say, on another quantum system (`the clock') has value `$T$'. Both systems evolve in the 
orthodox unitary (and non-interacting) manner with respect to the background time $t$. We 
work in the Heisenberg picture (with respect to $t$), and allow for continuous spectra. So 
the projector for the clock system for the value $T$ lying in the interval $[T_0 - \Delta 
T, T_0 + \Delta T]$ is
\be
P_{T_0}(t) : = \int_{T_0 - \Delta T}^{T_0 + \Delta T} \; dT \; \Sigma_k \; | T,k,t \rangle 
\langle T,k,t | \;\; ;
\label{P_T}
\ee
where: (i) the $t$ in the bra and ket signal the use of the Heisenberg picture; (ii) $k$ 
represents the eigenvalues of some operators that together with $\hat T$ form a complete 
set; and (iii) if the $k$ have a partly continuous spectrum the sum over $k$  should be 
appropriately replaced by an integral.

Similarly the projector for the measured system for the value $O$ lying in the interval 
$[O_0 - \Delta O, O_0 + \Delta O]$ is
\be
P_{O_0}(t) : = \int_{O_0 - \Delta O}^{O_0 + \Delta O} \; dO \; \Sigma_j \; | O,j,t \rangle 
\langle O,j,t | \; ,
\label{O_T}
\ee
where $j$ represents eigenvalues of operators that together with $\hat O$ form a complete 
set.

Then the (orthodox Born-rule) conditional probability that $\hat O$ takes value $O_0$ given 
that the clock indicates time $T_0$ is
\be
{\cal P}(O \in [O_0 - \Delta O, O_0 + \Delta O] | T \in [T_0 - \Delta T, T_0 + \Delta T]) =  
\frac{\lim_{\tau \rightarrow \infty} \;
\int_{- \tau}^{\tau} \; dt \; {\rm {Tr}}(P_{O_0}(t)P_{T_0}(t) \rho P_{T_0}(t))}
{\lim_{\tau \rightarrow \infty} \; \int_{- \tau}^{\tau} \; dt \; {\rm {Tr}}(P_{T_0}(t) 
\rho)}
\label{1stcondprob}
\ee
where: (i) $\rho$ is the (Heisenberg-picture, so $t$-constant) state of the combined 
system-plus-clock; (ii) the integrals over all $t$ on the right hand side reflect our 
ignorance about when in the background time $t$ the clock takes the value $T_0$; (iii) we 
envisage that $\Delta T$ is chosen much smaller than any characteristic time scales of the 
problem (in particular, the time-interval between successive measurements, if we were to 
consider multiple measurements); and (iv) we envisage that $\hat T$ is so chosen that it 
does not take  the same value twice (at least in the lifetime of the experiment).

Our aim now is to relate this in-principle expression to the usual expression for the 
Born-rule probability that $\hat O$ takes value $O_0$ at time $t$: which is, in the more 
familiar Schroedinger picture and an obvious notation:
\be
{\cal P}(O_0 | t ) = \frac{{\rm {Tr}}(P_{O_0}(0) \rho(t))}{{\rm {Tr}}(\rho(t))} \; .
\label{usual}
\ee

To this end, we now assume that the state is a product of the observed system and the 
clock, and that they never interact. So $\rho = \rho_{\rm{sys}} \otimes \rho_{\rm{cl}}$ and 
$U = U_{\rm{sys}} \otimes U_{\rm{cl}}$. Then the conditional probability eq. 
\ref{1stcondprob} is equal to:
\be
 \frac{\lim_{\tau \rightarrow \infty} \; 
 \int_{- \tau}^{\tau} \; dt \;
{\rm {Tr}}(U_{\rm{sys}}(t)^{\dagger} P_{O_0}(0) U_{\rm{sys}}(t) U_{\rm{cl}}(t)^{\dagger} 
P_{T_0}(0) U_{\rm{cl}}(t) \rho_{\rm{sys}}  \otimes \rho_{\rm{cl}} )
}
{\lim_{\tau \rightarrow \infty} \;
\int_{- \tau}^{\tau} \; dt \; {\rm {Tr}}(P_{T_0}(t) \rho_{\rm {cl}}){\rm {Tr}}( 
\rho_{\rm{sys}})} \; ;
\label{2ndcondprob}
\ee
which is equal to
\be
 \frac{\lim_{\tau \rightarrow \infty} \; 
 \int_{- \tau}^{\tau} \; dt \;
{\rm {Tr}}(U_{\rm{sys}}(t)^{\dagger} P_{O_0}(0) U_{\rm{sys}}(t) \rho_{\rm{sys}}){\rm {Tr}}( 
U_{\rm{cl}}(t)^{\dagger} P_{T_0}(0) U_{\rm{cl}}(t)  \rho_{\rm{cl}} )
}
{\lim_{\tau \rightarrow \infty} \;
\int_{- \tau}^{\tau} \; dt \; {\rm {Tr}}(P_{T_0}(t) \rho_{\rm {cl}}){\rm {Tr}}( 
\rho_{\rm{sys}})} \; .\footnote{I thank Gordon Fleming for pointing out that in eq.s 
\ref{1stcondprob} to \ref{3rdcondprob}, one needs in general a ratio of limits, so as to 
secure equality with the conditional probability on the left hand side of eq. 
\ref{1stcondprob}. In fact, Gambini et al (ibid.) write a limit of a ratio, i.e. one $\tau$ 
limit governing the ratio of two integrals: which is of course equal to the ratio of 
limits, provided both numerator and denominator have a limit. But I have followed Fleming's 
correction, to respect the concept of conditional probability justifying eq. 
\ref{1stcondprob}.}
\label{3rdcondprob}
\ee

We now define
\be
{\cal P}_t(T) := \frac{{\rm {Tr}}(P_{T_0}(0) U_{\rm{cl}}(t)  
\rho_{\rm{cl}}U_{\rm{cl}}(t)^{\dagger} )}
{\int_{- \infty}^{\infty} dt \; {\rm {Tr}}(P_{T_0}(t) \rho_{\rm {cl}})} \; ;
\label{REF7}
\ee
which obeys $\int_{- \infty}^{\infty} dt \; {\cal P}_t(T) = 1$, and represents a 
probability density of $t$ at given $T$; (not `of $T$ at given $t$', despite the notation).

We also define an ``effective'' density matrix (of the observed system) as a function of 
$T$, by
\be
\rho_{\rm {eff}}(T) : = \int_{- \infty}^{\infty} dt \; U_{\rm{sys}}(t)  
\rho_{\rm{sys}}U_{\rm{sys}}(t)^{\dagger} {\cal P}_t(T) \; .
\label{REF8}
\ee

With these definitions, it is easy to verify, by noting that
\be
{\rm {Tr}}(\rho(T)) = \int_{- \infty}^{\infty} dt \; {\cal P}_t(T) {\rm 
{Tr}}(\rho_{\rm{sys}}) = {\rm {Tr}}(\rho_{\rm{sys}}) \; ,
\ee
that the conditional probability eq. \ref{3rdcondprob} is equal to
\be
{\cal P}(O_0 | T ) := \frac{{\rm {Tr}}(P_{O_0}(0) \rho_{\rm {eff}}(T))}{{\rm 
{Tr}}(\rho_{\rm {eff}}(T))} \; :
\label{usualeffective}
\ee
which is of the familiar form, as in eq. \ref{usual}, except with the effective density 
matrix $\rho_{\rm {eff}}(T)$ substituted for the background-time density matrix $\rho(t)$. 
Thus the definition of  $\rho_{\rm {eff}}(T)$  makes eq. \ref{3rdcondprob}  take the 
familiar form eq. \ref{usual}.

To sum up: the statistics we gather for the quantity $\hat O$ at the clock time $T$ for a 
system prepared at $t = 0$ in $\rho_{\rm{sys}}$ are given by a familiar-looking formula, 
viz. eq. \ref{usualeffective}: but one which uses the density matrix $\rho_{\rm {eff}}(T)$ 
defined in eq. \ref{REF8}. Looking at that definition, we infer: the statistics are as if 
the system is drawn from an ensemble, whose components are (i) indexed by $t \in \mathR$, 
(ii) given by $U_{\rm{sys}}(t) \rho_{\rm{sys}} U_{\rm{sys}}(t)^{\dagger}$ and (iii) 
weighted with the probability density ${\cal P}_t(T)$ as given by eq. \ref{REF7}.

The physical idea is clear, and natural. Because the clock may be inaccurate, i.e. $T$ may 
be unequal to $t$, the statistics at a given $T$ are a mixture of the orthodox statistics 
associated with unitary evolutions to various times $t$, with the mixture's weights being 
given by ${\cal P}_t(T)$, the probability density at given $T$ of the component labelled 
$t$. Thus the lab technician might report: `I measure $O$ when the clock reads `$T$'. But 
the statistics of $O$ that I gather are as if each measurement draws the system from an 
urn, in the state $U_{\rm{sys}}(t) \rho_{\rm{sys}} U_{\rm{sys}}(t)^{\dagger}$ with a 
probability ${\cal P}_t(T)dt$'.

It is also clear that in terms of the clock-time $T$, evolution is not unitary: for 
$\rho_{\rm {eff}}(T)$ is a convex combination of density matrices, each of which is 
unitarily evolving with respect to $t$, but which are associated with {\em different} 
values of $t$. This {\em temporal decoherence} will be at the centre of the Montevideo 
interpretation.

The Montevideans write down the equation of motion for $\rho_{\rm {eff}}(T)$ in terms of 
$T$, on the assumption that the clock is ``good'', in the sense that the probability 
density ${\cal P}_t(T)$ takes the form
\be
{\cal P}_t(T) = f(T - t)
\ee
where $f$ achieves a maximum at $T = t$, and has narrow spread (i.e. decays rapidly, away 
from $T = t$).
As one would guess, the equation (derived by considering a Taylor expansion around $t$) is 
like the usual quantum Liouville equation for a Schroedinger-picture density matrix, but 
with additional terms. I will not need its exact form; (technically, it is a Lindblad 
equation, cf. e.g. eq. 14 of Gambini, Porto and Pullin (2007)). But I do need a solution of 
it. Namely: if one assumes that the the distribution $f$ is symmetric and its width grows 
linearly with respect to clock-time $T$, then the equation can be exactly solved; and if 
the system has a discrete energy spectrum, the  solution is
\be
\rho_{\rm {eff}}(T)_{nm} = [\rho_{\rm {eff}}(0)_{nm}] \exp(- i \omega_{nm}T)\exp(- \sigma 
\omega^2_{nm}T)
\label{simpledecay}
\ee
where: (i) $\omega_{nm} : = \omega_n - \omega_m \equiv (E_n - E_m)/{\hbar}$ is the 
difference of the frequencies for levels $n$ and $m$; and (ii) $\sigma$ is the (constant) 
rate of change, with respect to $T$, of the width of $f$. Thus the off-diagonal terms go to 
zero exponentially.

The first comment to make about the analysis just summarized is that it is clear and 
uncontentious. For it is an orthodox quantum-theoretic analysis, with a natural physical 
idea underlying its effective density matrix $\rho_{\rm {eff}}(T)$.  Besides, the 
Montevideans note that some other analyses, with different motivations than theirs, get the 
same or a similar result; and that the result successfully describes some real experiments 
with appropriately `bad' clocks $\hat T$. But as one would expect, it is unclear and even 
contentious how the mixture eq. \ref{REF8}, or its special case eq. \ref{simpledecay}, can 
contribute to solving the measurement problem---cf. the next Section.

\section{The meanings and roles of a mixture}\label{meaning1}
In this Section, I will argue that the mixture in the general form, eq. \ref{REF8},  does 
{\em not} fulfill the usual two roles for which a mixture is obtained, or hoped for, in 
approaches of the quantum measurement problem; while on the other hand, the special case 
eq. \ref{simpledecay} {\em does} fulfill the first role. I will do this by first defining 
the roles, together with two familiar approaches, viz. decoherence by the environment and 
dynamical reduction (Section \ref{two}). The situation for the Montevideans' mixtures will 
then be clear, in Section \ref{montmix}.

I stress that this contrast, as to fulfillment of the two roles, will not be an objection 
to Section \ref{montev1}'s analysis; nor even to the Montevideans' proposed use of it to 
solve the measurement problem. The basic reason why not is simply that the Montevideans do 
{\em not} ask temporal decoherence to fulfill the usual two roles for a mixture. Rather, 
they propose that it, especially the mixture's special form eq. \ref{simpledecay}, answers 
objections made to environment-induced decoherence---which they accept as playing a crucial 
part in solving the measurement problem. This proposal, appealing to considerations of 
quantum gravity, will be the topic of Section \ref{montev2}. But we will get a clearer view 
of the proposal, by first spelling out this contrast about the two roles.

\subsection{Two roles, two approaches and two debates}\label{two}
Thus consider the two  roles that a mixture, obtained in an analysis of quantum 
measurement, is usually taken as needing to fulfill. I give them obvious labels:\\
\indent (Basis): To define a preferred quantity, or basis, or set of nearly orthogonal 
states: and usually, one wants this quantity to be one that intuitively is definite at the 
end of measurement (e.g. a quantity such as the position of the centre of mass of a 
pointer) . This role---or the goal, or difficulty, of fulfilling it---is sometimes called 
`the preferred basis problem', or `the basis problem'.\\
\indent (Outcome): To secure that in each specific measurement trial, there is an 
individual outcome (represented mathematically by one component of the mixture). This 
role---or the goal, or difficulty, of fulfilling it---is sometimes called `the problem of 
outcomes'; or following d'Espagnat's jargon (1976, Section 6.6), `the problem of 
interpreting the mixture as proper (i.e. ignorance-interpretable) rather than improper 
(obtained by partial tracing)'; or following Bell's jargon, `the problem of replacing `and' 
by `or'. (The Montevideans use all three labels.)\\
\indent Broadly speaking, (Outcome) presupposes (Basis): only once a quantity or basis is 
preferred, can one raise the question of securing one of the quantity's values, or one of 
the basis' elements.

To illustrate what is at stake in fulfilling these roles, let us consider two familiar 
approaches to the measurement problem---which, as mentioned, will help us to `locate' the 
Montevideo interpretation. Again, I use labels:\\
\indent (Ein): Decoherence by the environment, as studied by the Heidelberg school of Zeh 
and co-authors, and the Los Alamos school of Zurek and co-authors; (cf. Giulini et al. 
(1996), Schlosshauer (2008). My label `Ein' abbreviates `einselection', a word invented by 
Zurek to convey `environment-induced superselection'.\\
\indent (DRP): The dynamical reduction programme of Ghirardi, Rimini, Weber, Pearle and 
others (begun, in large measure, by Ghirardi et al. (1986)). This programme models the 
collapse of the wave packet (Section \ref{CWP}) as a physical process.

Without going into details, I first report what I take to be the literature's consensus, 
which the Montevideans (and I) endorse. Namely: \\
\indent (Ein) fulfills (Basis): for decoherence selects a quantity like the position of the 
centre of mass of the apparatus pointer, in the sense that after the decoherence process, 
the state of the apparatus and measured system is a density matrix that is nearly diagonal 
in that quantity.\\
\indent But (Ein) does not fulfill (Outcome). This is expressed in various ways by various 
authors; cf. e.g. Giulini et al (1996, p. 37) and Leggett (2008). But the main point is 
that since the density matrix has been obtained by tracing over the environment, it 
represents an {\em improper}, not proper, mixture: there is no selection (even in some 
conceptual sense) of one of its components as against the others. Indeed, the  interference 
terms, that are characteristic of the {\em total} system (including the environment) being 
in a pure state (and thus symptomatic of the mixture being improper), can in principle be 
revealed by measuring a suitable quantity on the total system (a `global quantity').\\
\indent  (DRP), on the other hand, fulfills both (Basis) and (Outcome). It fulfills (Basis) 
by construction. For the revision of the Schroedinger equation explicitly exhibits the 
quantity (typically defined in terms of position and mass-density) with respect to which  
the final density matrix of the apparatus and measured system is nearly diagonal. And 
(Outcome) is fulfilled because this density matrix represents a proper, i.e. 
ignorance-interpretable, mixture: for it has not been obtained by tracing over the 
environment. Besides, (DRP)'s  scheme contains two other mathematical representations of 
the individual outcome of a specific measurement trial: viz. a much higher amplitude for 
that outcome than for the others, and an associated realization (also known as: trajectory)  
of the postulated stochastic process which determines the outcome in question; (whereas 
(Ein) has no such higher amplitude for a single outcome, and no such stochastic process).

Agreed, that is a hasty summary. There are various debates about whether (Ein) and (DRP) 
can succeed. To prepare for the Montevideans' proposals in Section \ref{montev2}, I need to 
note two such debates.

\subsubsection{Can we ignore small amplitudes?}\label{small} 
I just said that (Ein) and (DRP) each fulfill (Basis) by securing a final density matrix of 
the apparatus and measured system which is nearly diagonal in a quantity that intuitively 
one wants to be definite. But one common line of objection to (Ein) and (DRP) presses the 
question: is `{\em nearly}' good enough? The off-diagonal terms of the density matrix may 
have a tiny modulus, but they are physically real.\\
\indent Here we should distinguish three issues (and maybe more!). The first two apply, I  
believe, to both (Ein) and (DRP); the third, usually called the `problem of tails', applies 
only to (DRP). I only need to state them: I will not need to pursue them (thank 
goodness!).\\
\indent (i): {\em Revivals}: In some models,  after a very long time the off-diagonal terms 
become large: so that whatever definiteness of the preferred quantity we had won by the end 
of the measurement interaction turns out in the very long run to have been but a temporary 
victory. This is called {\em the problem of revivals} (or `recurrence of coherence', or 
`recoherence').  \\
\indent (ii): {\em Which quantity?}: One natural tactic for replying is to say that we 
should not be dogmatic {\em ab initio} about exactly which quantity is to be secured as 
definite, in order to solve the measurement problem. We should be flexible: if calculations 
suggest that a quantity different from the one we at first wanted definite, but in some 
sense close to it, is definite---then we should be satisfied with that. Thus suppose the 
density matrix at the end of the measurement interaction is nearly diagonal in the 
quantity, $\hat Q$ say, that we at first wanted definite, but is exactly diagonal in a 
quantity ${\hat Q}'$, with ${\hat Q}'$ and ${\hat Q}$ suitably close (say, in operator 
norm). Then  that is surely good enough: at least for early times---since we must admit 
this tactic does not answer the problem of revivals. \\
\indent (iii): {\em Tails}:  Finally, small-amplitude terms have been taken as an objection 
to (DRP)'s claim to have fulfilled (Outcome) by the final state having a much higher 
amplitude for some single outcome than for the others. Thus some people object that a 
possible outcome having a large, rather than  tiny, amplitude is no rationale for deeming 
it, rather than the other outcomes, to have occurred. Why not say instead that any outcome 
with a non-zero, albeit tiny, amplitude occurs? (This is called  {\em the problem of 
tails}.)

\subsubsection{Is vagueness a virtue or a vice?}\label{vague} 
Section \ref{small}'s debate about small amplitudes is part of a larger debate about 
whether a solution to the measurement problem must make an absolutely precise statement 
about which quantities on which systems have definite values (and with which 
probabilities).\\
\indent On the one hand, the tactic (ii) in Section \ref{small} suggests we can be, even 
should be, flexible.  Thus we require a solution to the measurement problem to vindicate 
the classical physical description (`the appearance of a classical world', as it is 
sometimes put)---but only approximately. Only some subset of quantities should have 
definite values. And maybe that subset should only be specified contextually, even vaguely. 
And maybe the values should only be definite within some margins of error, even vague ones. 
Clearly, (Ein)  secures this sort of flexibility. For the selection of the preferred 
quantity is made by a dynamical process---whose definition can be legitimately varied in 
several ways. Three examples: the definitions of:\\
\indent (a) the system-environment boundary (and therefore of what exactly is the 
interaction hamiltonian), and of \\
\indent (b) the time at which the interaction ends, and of \\
\indent (c) what counts as a state being 'sharply peaked' for a quantity,\\
can all be varied. Of course, item (c) echoes Section \ref{small}'s debate about small 
amplitudes. And (DRP) shares with (Ein) the flexibility about defining (b) and (c); but not 
(a), since it makes no system-environment split.\\
\indent  On the other hand, there is a strong impulse to require precision; for (at least) 
two reasons. First: over the eight decades since the measurement problem was first 
articulated, many proposals for how to solve it have been vague in ways that undermined the 
proposal: the vagueness amounted to hand-waving (as the writings of Bell often lamented). 
Second: most philosophers construe vagueness as semantic indecision. That is: our words and 
concepts being vague is a matter of our not having decided on a precise boundary between 
instances and non-instances---and not a matter of reality itself being vague (a notion 
which may well not be coherent). Most philosophers also take mathematics to be precise, 
i.e. not vague. They also take physics to be in large part precise, often because it is 
expressed in mathematical language. And where it is not precise (as in much heuristic 
physics), they take it nevertheless to aspire to precision. So they would naturally require 
quantum theory, in purporting to be a fundamental  theory, to give, or aspire to give, a 
precise i.e. non-vague solution to its measurement problem.\\
\indent I will return to this debate, and my own take on it, in Section \ref{montev2}.

So much by way of summarizing, partly in preparation for Section \ref{montev2}: the two 
usual roles for a mixture, how the two approaches (Ein) and (DRP) treat those roles, and 
the debates about their treatments. In the light of those debates, I should perhaps weaken 
what I called `the literature's consensus' about these approaches' fulfilling the roles, 
and speak only of {\em aims}. That is: a more cautious summary would be to say just that 
(Ein) {\em aims} to fulfill (Basis), but does not aim to fulfill (Outcome); while  (DRP) 
aims to fulfill both (Basis) and (Outcome).

In any case, this summary is enough for my purposes. For we can now check whether the 
Montevideo interpretation's temporal decoherence mixtures, eq. \ref{REF8} and eq. 
\ref{simpledecay}, fulfill the roles (Basis) and (Outcome).

\subsection{Montevideo's mixtures}\label{montmix}
I begin with the first, general form, eq. \ref{REF8}. Clearly, it does not fulfill the 
roles (Basis) and (Outcome). As to (Basis), eq. \ref{REF8} does not suggest an orthogonal, 
or nearly orthogonal, decomposition of $\rho(T)$ even if the initial state 
$\rho_{\rm{sys}}$ is pure (given as $| \psi \rangle \langle \psi |$). For of course typical 
components labelled by different values of $t$, $U_{\rm{sys}}(t_1) \rho_{\rm{sys}} 
U_{\rm{sys}}(t_1)^{\dagger}$ and $U_{\rm{sys}}(t_2) \rho_{\rm{sys}} 
U_{\rm{sys}}(t_2)^{\dagger}$, will not be orthogonal. Looking ahead for a moment: this 
aspect will remain when, in Section \ref{montev2}, we `throw away the ladder' of assuming a 
background time $t$.

As to (Outcome), the situation is a little subtle. I said that (Outcome) presupposes 
(Basis): so since (Basis) is not fulfilled, (Outcome) can hardly be. But on the other hand, 
I stress that since we are at present assuming that there is a background time $t$, each 
individual system measured at clock time $T$ has in fact a time $t$: i.e. has in fact 
persisted from the initial time $t = 0$ to a time $t$, at which it is measured for $O$ and 
the clock says `$T$' (with maybe $T \neq t$!). In short: though the value of $t$ does not 
deserve the name `outcome' (which presumably should here be reserved for entities 
correlated with values of the quantities $\hat O$ and $\hat T$), this value is perfectly 
definite in each specific measurement trial. But as one would guess: this `definiteness of 
$t$' will disappear when, in Section \ref{montev2}, we `throw  away the ladder' of assuming 
a background time $t$.

I turn to the special case, eq. \ref{simpledecay}. This is nearly diagonal in a basis that 
is of central physical significance, the energy basis. So the situation is obviously 
similar to that which occurs in decoherence by the environment. That is: eq. 
\ref{simpledecay} fulfills the first role, (Basis). It selects energy as the preferred 
quantity: of course, {\em modulo} the debate in Section \ref{small}, about whether we may 
ignore small amplitudes---in particular the problem of revivals. And eq. \ref{simpledecay} 
does not fulfill, and does not aim to fulfill, the second role, (Outcome): no component of 
the mixture is selected.

In the light of this discussion, it is natural to ask how if at all the Montevideo 
interpretation will fulfill (Basis) and (Outcome). As I see matters, the short answer---to 
be expanded in Sections \ref{montev2} and \ref{enfin}--- is as follows.

\indent As to (Basis): The interpretation appeals to a combination of (Ein) and temporal 
decoherence, in the specific form of eq. \ref{simpledecay}. More specifically, temporal 
decoherence is proposed as solving the problem of revivals that besets (Ein)'s fulfillment 
of (Basis); (cf. Section \ref{small}). The idea will be that temporal decoherence makes 
revivals unobservable in principle, not merely in practice.

\indent As to (Outcome): The situation is less clear to me. To explain this, I should 
distinguish two aspects of (Outcome) which I so far have not needed to separate:\\
\indent \indent (i) justifying the ascription of a proper (i.e. ignorance-interpretable) 
mixture; \\
\indent \indent (ii) justifying the ascription to an individual system of a single 
component of such a mixture.

As I see matters, the Montevideans succeed in (i). They give details about how temporal 
decoherence, especially eq. \ref{simpledecay}, answers the usual `global quantity' 
objection that, as mentioned in Section \ref{two}, prevents (Ein) fulfilling (Outcome). 
Their idea will be similar to their idea just above, for (Basis): temporal decoherence 
makes the global quantity unmeasurable in principle, not merely in practice. 

But what about (ii): securing an individual outcome in each specific measurement trial? 
Here, I think the situation is less clear. I myself think their interpretation fits best in 
an Everettian picture, for reasons to do with the vagueness debate of Section \ref{vague}. 
But some of their remarks suggest they disagree. But as I mentioned, Sections \ref{montev2} 
and \ref{enfin} will expand on this.

\section{Throwing away the ladder}\label{montev2}
The title refers to Wittgenstein's metaphor of throwing away a ladder once one has climbed 
it to gain an understanding. Here, the ladder is the background time $t$: and nowadays the 
various research programmes seeking a quantum theory of gravity make rival proposals for 
how to throw away this ladder. As mentioned in Section 2.1.3, the Montevideans work in one such programme, viz. quantum general relativity; but the 
Montevideo interpretation does not depend on its details. We can state their interpretation 
as five proposals. The first four proposals are treated in Section \ref{fourproposals}. The 
fifth proposal is separated out, in Section \ref{fifthproposal}, since it leads in to  
Section \ref{enfin}'s concluding suggestion that the Montevideo interpretation is best seen 
as Everettian.

\subsection{Four proposals}\label{fourproposals}
 The Montevideans  propose that:\\
\indent (i): Some heuristic arguments about black holes suggest that there are fundamental 
limits to how accurate a quantum clock can be, so that the temporal decoherence of Section 
\ref{montev1} cannot be completely eliminated. \\
\indent (ii): The equation of motion in terms of a clock-quantity $T$ for a clock that is 
as accurate as possible should be taken as fundamental. This is the `throwing away of the 
ladder' of background time $t$. So the ineliminable decoherence is now a fundamental or 
principled type of decoherence.\\
\indent (iii): This fundamental decoherence, together with environmental decoherence, 
solves the quantum measurement problem, along the lines given in Section \ref{montmix}.\\
\indent (iv): Besides, although the arguments in (i) vary somewhat in the accuracy limits 
they suggest, the ensuing solution to the measurement problem is robust: it holds good if 
one adopts the other  accuracy limits.

I will present (i)-(iv), in order (Sections \ref{heur} to \ref{robust}). But my discussion 
will be much briefer than in Sections \ref{montev1} and \ref{meaning1}. This reflects not 
just lack of space: also, the authors' arguments are more heuristic; and so my own grip on 
them is more tentative. These technicalities will lead in to the Montevideans' final 
interpretative proposal, about under what exact conditions `the wave-packet collapses' 
(Section \ref{fifthproposal})---which will prompt me to suggest in Section \ref{enfin} that 
the interpretation fits best in an overall Everettian approach.

\subsubsection{Heuristic arguments}\label{heur}
As I announced in (ii) of Section \ref{clocks}, the Montevideans endorse arguments by 
various authors, especially Salecker and Wigner, and Ng and van Dam, suggesting  
fundamental limits on quantum clocks. Thus Salecker and Wigner argued that the mass of a 
clock limited its accuracy: specifically,  if $T$ is the time-interval to be measured by a 
clock of mass $M$, then there must be an  uncertainty $\delta T$ proportional to 
$\surd{T/M}$ (1957, pp. 572-574; cf. also Wigner  (1957, pp. 260-61)). This suggests that 
one can attain arbitrary accuracy by a sufficiently massive clock. 

But Ng and van Dam (1994) argued that in general the mass needs to be confined; and that 
therefore the attainable mass, and so accuracy, is limited by the onset of the formation of 
a black hole. According to their analysis:\\
\indent (a) if $T$ is again the time-interval to be measured, the maximum attainable mass  
is given by the mass of a black hole whose Schwarzschild radius is the distance light would 
travel in $T$; and \\
\indent (b) the corresponding maximum attainable accuracy is given by an uncertainty 
$\delta T$ proportional to $T_{\rm{Planck}}^{2/3}T^{1/3}$,  where $T_{\rm{Planck}} = 
10^{-44}$ seconds is the Planck time. So for typical laboratory times of a few hours, 
$\delta T \sim 10^{16}T_{\rm{Planck}} \sim 10^{-28}$ seconds. This is not now 
experimentally measurable, but in principle it could be observed.

This analysis remains heuristic, indeed apparently controversial; (cf. the exchange between 
Baez and Olson (2002) and Ng and van Dam (2003)). But the Montevideans endorse this kind of 
analysis: taking heart from the fact (Section \ref{robust}) that their solution to the 
measurement problem seems insensitive to what are the exact limits on clocks' accuracy.

\subsubsection{Fundamental decoherence}\label{fund}
Prompted by the arguments in Section \ref{heur}, the Montevideans propose to dispense with 
the time $t$ that formed the background of Section \ref{montev1}'s analysis: the 
fundamental equation of motion is to be their equation of motion in terms of  $T$, the 
clock-quantity for a maximally accurate clock. 

Or rather, to be precise: they take the arguments in Section \ref{heur} to suggest that the 
unknown theory of quantum gravity, whatever its exact form, will yield their equation of 
motion using this $T$, as the effective equation of motion governing a regime, whose exact 
definition we of course cannot give (being ignorant of the quantum gravity theory!)---but 
which we expect to include all the successful applications of quantum theory, both 
non-relativistic and relativistic, ``away from'' quantum gravity effects such as might 
occur in black holes.\footnote{Like almost all the literature on decoherence, the 
Montevideans' analyses look non-relativistic; but one can imagine generalizing them to get 
an explicit Lorentz-invariance. Recall Section \ref{grav?}'s contrast between general 
relativity's treating time as dynamical, and quantum theory's treating it as `background', 
as in the cousin classical, Newtonian or special-relativistic, theories.}

\indent This proposal implies that attention focusses on Section \ref{montev1}'s exact 
solution to their equation, for a system with discrete energy levels,   eq. 
\ref{simpledecay}, but now with the width  $\sigma(T)$ given as 
$\left(\frac{T_{\rm{Planck}}}{T}\right)^{1/3}.T_{\rm{Planck}}$, i.e.
 \be
\rho_{\rm {eff}}(T)_{nm} = [\rho_{\rm {eff}}(0)_{nm}] \exp(- \omega_{nm}T)\exp(- 
\omega^2_{nm}T_{\rm{Planck}}^{4/3}T^{2/3}) \; .
\label{simpledecay2}
\ee

\subsubsection{Fundamental decoherence allied with environmental decoherence}\label{allied}
We can now sketch how eq. \ref{simpledecay2} is meant to supplement the process of 
environmental decoherence (called (Ein) in Section \ref{two}), so as to solve the 
measurement problem. As I announced in Section \ref{montmix}, the idea is  that this 
fundamental temporal decoherence kills the off-diagonal terms of the density matrix of the 
macroscopic apparatus (together with the measured system). In particular, this solves:\\
 \indent (i): the problem of revivals, i.e. the fact that in some models {\em without} 
temporal decoherence, the off-diagonal terms can in principle, in the very long run, become 
large. (In terms of Section \ref{small}'s debate  about small amplitudes: this problem 
confronts (Ein)'s fulfillment of the role (Basis).)\\
 \indent (ii): the problem of global quantities, i.e. the fact that in some models {\em 
without} temporal decoherence, one can in principle measure a quantity whose statistics 
reveal the total system, including the environment, to be in a pure (entangled) state. (In 
terms of Section \ref{two}: this problem confronts (Ein)'s fulfillment of the role 
(Outcome).)

Here is a bit more detail about (i); (I set aside (ii), for which the discusssion is 
similar). The general idea is that without temporal decoherence, an off-diagonal term is a 
product of $N$ complex numbers of unit modulus $\exp (i (\theta_k t + a_k)), k = 1,..., N$, 
where $N$ represents the number of degrees of freedom of the apparatus (and measured 
system), and so is typically very large, and the phases $a_k$ are randomly distributed. 
Because $N$ is large, the common period of oscillation for the terms is very long, maybe 
even comparable with the age of the universe---yet nevertheless finite. But when we include  
temporal decoherence in the calculation, there is an extra exponential suppression. Thus 
one detailed analysis (Gambini, Garcia Pintos and Pullin, 2010, Sections IV-V) studies a 
spin model of Zurek's, and a more realistic modification of it. Owing to the temporal 
decoherence, the off-diagonal terms get multiplied by an $N$-fold product of exponentials 
whose exponents are proportional to $T^{2/3}$. Putting in the figures for the models, this 
gives unobservably small tails, and so no revivals even in very long times, even for $N$ 
between 100 and 1000.

\subsubsection{Robustness}\label{robust}
In Section \ref{heur}, I reported that heuristic arguments suggest that the most accurate 
clock has an uncertainty $\delta T$ proportional to $T_{\rm{Planck}}^{2/3}T^{1/3}$. Here we 
should note that the Montevideans admit that some arguments suggest instead that $\delta T 
\sim T_{\rm{Planck}}^{1/2}T^{1/2}$.\\
\indent But this makes essentially no difference to their conclusions, or even to their 
quantitative results in Section \ref{allied}, about their proposed solution to 
environmental decoherence's problems of revivals and global quantities. For provided that 
the uncertainty in time goes as a positive power of $T$, one loses coherence. And 
quantitatively, as regards the solution to the problem of revivals, reported in Section 
\ref{allied}: if one were to postulate more cautiously that the uncertainty goes like a 
small positive power, i.e. $\delta T \sim T_{\rm{Planck}}^{1 - \varepsilon}T^{\varepsilon}$ 
for some small $\varepsilon$, then the only effect would be that the value of $N$ required 
to prevent revivals even in very long times would be raised from $N \sim 100$ to about $N 
/(3\varepsilon)$.

\subsection{The fifth proposal: about exactly when is there an outcome}\label{fifthproposal}
So much by way of reporting the four proposals. At this point, one naturally wants to press 
the following straightforward point:
\begin{quote}
Temporal decoherence involves exponentials: its suppression of interference is by dictating 
that exponential factors should multiply the terms that in environmental decoherence remain  
troublesome. So {\em the trouble persists}. That is: the terms remain, generically, 
non-zero for arbitrarily large times etc. So, presumably, the wave-function does not in 
fact collapse. More precisely: these four proposals have not secured that in each specific 
measurement trial, there is an individual outcome. In short: we have not yet passed from 
Bell's `and' to his `or'.' To put the point in Section \ref{two}'s jargon about the roles 
that a mixture might fulfill: the four proposals have not fulfilled the  role (Outcome).)
\end{quote}
The Montevideans are of course aware of this point. In answer, they make the following 
interpretative proposal, which completes their exposition. (At least: it completes it, in 
the order of exposition I have adopted in this paper.)

They say: there is transition from `and' to `or'---an 
individual outcome, a selection of a single component of the mixture, an event---exactly 
when it becomes in principle {\em undecidable} (thanks to fundamental temporal decoherence) 
whether or not the evolution of the total system is unitary. Here, philosophers will find 
the word `undecidable'  reminiscent of logic and computability theory; but these 
connotations are not intended. All that is intended is just that no measurement of any 
quantity, even in principle, can ascertain whether the evolution is unitary. Thus for a 
statement of this proposal, cf. Gambini, Garcia Pintos and Pullin (2010, Section III and 
Appendix 2.A), Gambini and Pullin (2009: replies to questions 10 and 11). They develop the 
proposal in more detail, for their modification of Zurek's spin model (mentioned at the end 
of Section \ref{allied}), in Gambini, Garcia Pintos and Pullin (2011, Sections 3-5). I will 
not enter into details, but just make two points: the second leading in to Section 
\ref{enfin}. 

First: this last discussion builds on some remarkable in-principle limitations on measuring 
spins proposed by Kofler and Brukner (2010). Second: nevertheless, the undecidability 
proposal about exactly when is the collapse, remains, in the Montevideans' publications so 
far, tied to specific models---prompting the question `What is their exact and general 
proposal?'

\section{Everett {\em en fin}}\label{enfin}
Here, I suggest, we see at last that the Montevideo interpretation fits best in an 
Everettian approach, in which a `branching'---in the above jargons: a transition from `and' 
to `or', an outcome, a selection, an event---is effective and approximate. And this returns 
us (as I promised) to the debate in Section \ref{vague}, about whether vagueness is a 
virtue or a vice.

For of course, in the usual Hilbert space formalism, any pure state and any mixed state are 
distinguished by the statistics, according to the Born rule, of some self-adjoint operator 
or other. So a rigorous statement of undecidability in the Montevideans' sense would 
involve, just as a matter of logic, a rigorous statement of what is the proposed 
restriction on quantities: which the Montevideans have not so far given.\footnote{Though 
this sort of rigorous statement is often made by those adopting the algebraic approach to 
quantum theory, such statements can of course be made using the Hilbert space formalism: 
which is the formalism the Montevideans adopt.} 

Thus as I see matters, the Montevideans face a strategic choice in developing their 
interpretation:\\
\indent (1): If they feel obliged to make such a statement: then their interpretation, 
though undoubtedly already of great interest, is not yet complete. Of course, 
incompleteness is no discredit: especially since one might think we should wait for the 
heuristic gravitational arguments in Section \ref{heur} to be made rigorous and-or to be 
extended, so as to fill out the arguments of Sections \ref{fund} and \ref{fifthproposal}.\\
\indent (2): If they do not feel so obliged: then this should be---I suggest---because they 
view such vagueness as a virtue, not a vice.

And as discussed in Section \ref{vague}, such vagueness is indeed a virtue on an Everettian 
approach, according to which the `branching' is phenomenological or effective in 
physicists' sense. That is: branching is an approximation whose value does not depend on 
its being precisely defined---indeed, the value depends on its {\em not} being precisely 
defined. We saw there how environmental decoherence  secured the desired flexibility about 
which quantities have definite values, by its flexibility in the definitions of:\\
\indent (a) the system-environment boundary, \\
\indent (b) the time at which the decohering interaction ends, and \\
\indent (c) what counts as a state being `sharply peaked'. \\
We can now add that the Montevideans' temporal decoherence has similar flexibility about 
(b) and (c). Thus I suggest that if the Montevideans choose (2), then their interpretation 
fits best with an Everettian picture of `effective branching'.

\indent Think of how valuable we find it to describe the region where a river meets the sea 
in terms of a delta composed of islands and channels. We do not need to decide what 
precisely counts as a channel, or a fork in the waterway. Indeed the description is more 
valuable precisely because we do not decide: to give precise definitions would be a waste 
of time, and would make our language and thought prolix and inflexible, and so less 
adaptable to the demands of a passing occasion. The helmsman who asks `Shall I take the 
left or the right channel?' would be annoyed to be told `It depends on how one defines 
`main channel''.




But in closing, I stress that my overall aim here has been, not to object to the 
interpretation---but to praise it! It has the merit of bringing specific, albeit heuristic, 
considerations of quantum gravity to bear on the measurement problem. So I invite the 
reader to explore it. \\

{\em Acknowledgements}:--- For comments on a previous (2012) version, I am very grateful to: Gordon Fleming, Rodolfo Gambini, Jorge Pullin, David Sloan, James Yearsley, David Wallace and 
Chris W\"{u}thrich; and to audiences in Dubrovnik, Croatia, and Cambridge and Oxford, UK. My thanks also to the editor and two referees of their comments. This work is also supported in part by a grant from the Templeton World Charity Foundation, whose support is gratefully acknowledged. \\

\section{References}
 
Baez, J. and Olson, S (2002), `Uncertainty in measurements of distance', {\em Classical and 
Quantum Gravity} {\bf 19} L121; arxiv: gr-qc/0201030v1

Bell, J. (1986), 'Six possible worlds of quantum mechanics', {\em Proceedings of the Nobel 
Symposium 65} (Stockholm August 1986); reprinted in Bell (1987), page references to 
reprint.

Bell, J. (1987), {\em Speakable and Unspeakable in Quantum Mechanics}, Cambridge: Cambridge 
University Press; second edition 2004, with an introduction by Alain Aspect.

Busch, P. (1990),  `On the energy-time uncertainty relation: Part I: Dynamical time and 
time indeterminacy', {\em Foundations of Physics} {\bf 20}, pp. 1-32.

Busch, P. (1990a),  `On the energy-time uncertainty relation: Part II: Pragmatic time 
versus energy indeterminacy', {\em Foundations of Physics} {\bf 20}, pp.  33-43.

Busch, P. (2008), `The time-energy uncertainty relation', Chapter 3 of {\em Time in Quantum 
Mechanics}, ed. G. Muga et al., Springer Verlag (second edition, 2008); quant-ph/0105049.

Butterfield, J. (2002), `Some Worlds of Quantum Theory ', in R.Russell, J. Polkinghorne et 
al (ed.), {\em Quantum Mechanics (Scientific Perspectives on Divine Action vol 5)}, Vatican 
Observatory Publications, 2002; pp. 111-140. Available online at: 
arXiv.org/quant-ph/0105052; and at http://philsci-archive.pitt.edu/archive/00000204.

Butterfield, J. (2012), `On time in quantum physics', in {\em The Blackwell Companion to 
Time}, edited by A. Bardon and H. Dyke; Oxford: Blackwell; pp. 220-241; available at 
http://philsci-archive.pitt.edu/9287/

Butterfield, J. and Isham C. (2001), `The philosophical challenge of quantum gravity', in 
C. Callender and N. Huggett (eds.), {\em Physics Meets Philosophy at the Planck Scale}, 
Cambridge University Press, pp. 33-89; gr-qc/9903072;\\ 
http://philsci-archive.pitt.edu/1915/

Dennett, D. (1991), `Real patterns', {\em Journal of Philosophy, 88}, pp.  27-51.

d'Espagnat, B. (1976), {\em Conceptual Foundations of Quantum Mechanics}, Reading, MA: W. 
A. Benjamin

Deutsch, D. (2010), `Apart from universes', in Saunders et al (eds.) (2010), pp. 542-552.

Gambini, R., Garcia Pintos, L. and Pullin J. (2010), `Undecidability and the problem of 
outcomes in quantum measurements', {\em Foundations of Physics} {\bf 40}, pp. 93-115; 
arXiv:0905.4222

Gambini, R., Garcia Pintos, L. and Pullin J. (2011), `Undecidability as solution to the 
problem of measurement: fundamental citerion for the production of events', {\em 
International Journal of Modern Physics} {\bf D20}, pp. 909-918; arXiv:1009.3817

Gambini, R., Garcia Pintos, L. and Pullin J. (2011a), `An axiomatic formulation of the 
Montevideo interpretation of quantum mechanics', {\em Studies in the History and Philosophy 
of Modern Physics} {\bf 42}, pp. 256-263; arXiv:1002.4209v2

Gambini, R.,  Porto R. and Pullin J. (2006), `Fundamental spatiotemporal decoherence: a key 
to solving the conceptual problems of black holes, cosmology and quantum mechanics', {\em 
International Journal of Modern Physics} {\bf D15}, pp. 2181-2186; arXiv:gr-qc/0611148

Gambini, R.,  Porto R. and Pullin J. (2007), Fundamental decoherence from quantum gravity: 
a pedagogical review', {\em General Relativity and Gravitation} {\bf 39}, pp. 1143-1156; 
arXiv:gr-qc/0603090

Gambini, R.,  Porto R. and Pullin J. (2008), `Loss of entanglement in quantum mechanics due 
to the use of realistic measuring rods', {\em Physics  Letters A} {\bf 372}, pp. 1213-1218; 
arXiv:0708.2935

Gambini, R.,  Porto R., Pullin J. and Torterolo S. (2009), `Conditional probabilities with 
Dirac observables and the problem of time in quantum gravity', {\em Physical Review } {\bf 
D79}, pp. 041501; arXiv:0809.4235v1

Gambini, R. and  Pullin J. (2007), `Relational physics with real rods and clocks and the 
measurement problem of quantum mechanics',{\em Foundations of Physics} {\bf 37}, pp. 
1074-1092; arXiv: quant-ph/0608243

Gambini, R. and  Pullin J. (2009), `The Montevideo interpretation of quantum mechanics: 
frequently asked questions' {\em Journal of Physics Conference Series, proceedings of the 
DICE 2008 Castiglioncello meeting}, {\bf 174}, 012003; arXiv:0905.4402.

Gambini, R. and  Pullin J. (2009a), `Free will, undecidability and the problem of time in 
quantum gravity' {\em Essay for FQXi}; arXiv:0903.1859 [quant-ph].

Ghirardi, G., Rimini, A. and Weber, T.  (1986), `Unified dynamics of microscopic and 
macroscopic systems', {\em Physical Review} {\bf D34}, pp. 470-491.

Giulini, D., Joos, E., Kiefer, C., Kupsch J., Stamatecu I. and Zeh. H (1986), {\em Decoherence and the Appearance of a Classical World in Quantum Theory}, Berlin: Springer.

Kent, A. (2010), `One world or many?' in Saunders, Barrett, Kent and Wallace (Eds.) (2010), pp. 307-354.

Kofler, J. and  Brukner C. (2010), `Are there fundamental limits for observing quantum 
phenomena from within quantum theory?'; arXiv: 1009.2654

Leggett, A. (2008), `Realism and the physical world', {\em Reports on Progress in Physics}, 
{\bf 71}, 022001; doi:10.1088/0034-4885/71/2/022001

Messiah, A. (1965), {\em Quantum Mechanics}, volume 1, North Holland.

Muga, J., Sala Mayato, R. and Egusquiza I. (eds.) (2008), {\em Time in Quantum Mechanics}, 
Berlin; Springer, two volumes.

Ng, Y. and van Dam, H. (1994), {\em Modern Physics Letters} {\bf A9}, pp. 335-340.

Ng, Y. and van Dam, H. (2003), `Comment on `Uncertainty in measurements of distance'', {\em 
Classical and Quantum Gravity}, {\bf 20}, p. 393; arxiv: gr-qc/0209021.

Rovelli, C. (2006), `Quantum gravity', in J. Butterfield and J. Earman (eds.), {\em 
Philosophy of Physics}, in the series {\em The Handbook of Philosophy of Science}, 
Amsterdam: Elsevier, pp. 1287-1330.

Salecker, H. and Wigner, E. (1957), `Quantum limitations of the measurement of spacetime 
distances', {\em Physical Review} {\bf 109}, pp. 571-577.

Saunders S.,  Barrett J.,  Kent A. and  Wallace D. (eds.) (2010), {\em Many Worlds? 
Everett, Quantum Theory and Reality}, Oxford: Oxford University Press.

Schlosshauer, M. (2008), {\em Decoherence and the quantum to classical transition}, Berlin: 
Springer.

Wallace, D. (2001), `Worlds in the Everett interpretation', {\em Studies in the History and 
Philosophy of Modern Physics} {\bf 33}, pp. 637-661.

Wallace D. (2012), {\em The Emergent Multiverse}, Oxford: Oxford University Press.

Wallace, D. (2012a), `Decoherence and its role in the modern measurement problem', 
{\em Philosophical Transactions of the Royal Society of London} {\bf A 370}, pp. 4576-4593;
http://uk.arxiv.org/abs/1111.2187

Wigner, E. (1957), `Relativistic invariance and quantum phenomena', {\em Reviews of Modern 
Physics} {\bf 29}, pp. 251-268.

Yearsley, J. (2011), `Aspects of time in quantum theory', arxiv: 1110.5790

\end{document}